\documentclass[10pt]{revtex4}
\pdfoutput=1

\usepackage[english]{babel}
\usepackage{graphicx}
\usepackage{amssymb}
\usepackage{amsmath}

\def\laq{~\raise 0.4ex\hbox{$<$}\kern -0.8em\lower 0.62ex\hbox{$\sim$}~}
\def\gaq{~\raise 0.4ex\hbox{$>$}\kern -0.7em\lower 0.62ex\hbox{$\sim$}~}

\def\beq{\begin{equation}}
\def\eeq{\end{equation}}
\def\bea{\begin{eqnarray}}
\def\eea{\end{eqnarray}}
\def\bean{\begin{eqnarray*}}
\def\eean{\end{eqnarray*}}

\begin{document}

\begin{center}
{\huge \bf  Anisotropic expansion of the Universe and generation of quantum interference in light propagation}\\

\vspace{10mm}

\vspace{3mm}
{G. Fanizza$^{1,2,3}$ and L. Tedesco$^{1,2}$} \\
\vspace{6mm}
{\sl ${ }^1$Istituto Nazionale di Fisica Nucleare, Sezione di Bari, Bari, Italy}\\
\vspace{3mm}
{\sl ${ }^2$Dipartimento di Fisica, Universit\`a di Bari, Via G. Amendola 173\\70126 Bari, Italy}\\
\vspace{3mm}
{\sl${ }^3$Universite de Geneve, Departement de Physique Theorique,\\ 24 quai Ernest-Ansermet, CH-1211 Geneve 4, Switzerland}
\end{center}
\begin{abstract}
We investigate the  electrodynamic in a Bianchi type I cosmological model. This scenario reveals the possibility that photons, during their traveling, can make quantum interference. This effect is only due to the presence of two different axes of expansion in the cosmic evolution. In other word, it is possible to conclude that a purely metrical - or, equivalently, gravitational - phenomenon gives rise up to a quantum effect that manifests itself in the light propagation.
\end{abstract}

\keywords{anisotropy, Bianchi I, quantum interference}

\maketitle
\section{Introduction}
The present Universe is homogeneous and isotropic on large scales. The most important evidence for homogeneity and isotropy is the highly smooth and uniform cosmic microwave background. On the other hand the (apparent) homogeneity of matter distribution may be at scales  $> 100 {h^{-1}} Mpc$. The inhomogeneity and/or the anisotropy influence the average evolution as it was put in evedence in the first work \cite{shirokov} some years ago and discussed in detail in \cite{ellis} under the name "fitting problem". If also at a large scale the Universe was almost isotropic, the studies of the effects of an anisotropic Universe in the primordial time makes the Bianchi type I model a very interesting alternative for study. The Universe is locally far from homogeneity and isotropy due the non-linear structures at late time.Very interesting works indicate possible deviations from the homogeneity and isotropy obtained when we consider angular distribution of the fine structure constant in the range of the redshift  $0.2223 < z < 4.1798$ when we measure the quasar absorption line spectra with the so called many multiplet method \cite {{mariano}, {mariano2}}. A very important consequence of the so called large-angles cosmic microwave background anomalies may be interpreted as a manifestation of a prefered direction in the Universe    \cite{{deoliveira}, {schwarz}, {land}, {Campanelli:2006vb}, {Campanelli:2007qn}, {abramo1}, {copi}, {wiaux}, {akerman}, {samal}, {naselski}, {copi2}}. A very interesting phenomena appear when we consider the equation of electrodynamics in the context of anisotropic expansion of the Universe \cite{Ciarcelluti:2012pc}. On the other hand interesting properties has been already studied as regards the particles created from the vacuum by curved space-time of an expanding spatially-flat Friedmann Lemaitre Robertson Walker Universe \cite{{Parker:1968mv},{Parker:2012at}}.

The goal of the present paper is to study Maxwell equations in anisotropic axisymmetric Bianchi type I cosmological model. In particular in this Letter  we study a very interesting effect due to presence of two axes of expansion  in an ellipsoidal Universe.
\section{Electrodynamic in a Bianchi I metric}
Let us begin by briefly discussing the quantization of electromagnetic field in a Bianchi type I space-time. The spatially homogeneous and anisotropic Bianchi type I model is descripted by the line element
\begin{equation}
ds^2=dt^2-a^2(t)\left[ dx^2+dy^2\right]-b^2(t)dz^2
\end{equation}
where scale factor $a$ and $b$ are functions of cosmic time $t$ only, 
$z$ is the direction of anisotropy and $x$-$y$ plane contains a planar symmetry.
In such a way, we are able to derive the equations for electrodynamic, starting from Maxwell equations in a curved space-time so that, requiring the generalized gauge $\frac{\partial_1A_1}{a^2(t)}+\frac{\partial_2A_2}{a^2(t)}+\frac{\partial_3A_3}{b^2(t)}=0$, where $A_\mu$ is the quadripotential for the electromagnetic field. We consider the simple case in which $A_0=0$, the equations become:
\begin{equation}
\sum_l\frac{\partial_lA_l}{a^2_l}\frac{\dot a_l}{a_l}=0
\end{equation}
\begin{equation}
\sum_{j=1}^3 
\frac{\partial^2 A_i(t,\vec r)}{\partial r_j^2}-\partial_0^2 A_i(t,\vec r)-\partial_0\ln{\frac{\bar a^3}{a_i^2}}\,\partial_0 A_i(t,\vec r)=0
\end{equation}
where $dr_j = a_j(t)\,dx_j$, $\bar a^3=a^2 b$, $a_l=a$ for $l=1,2$ and $a_l=b$ for $l=3$.
\\
At this step, let's define the normal modes $A_{ik}$ in the following way:
\begin{equation}
A_i (t,\vec r)=\int\frac{d^3k}{(2\pi)^3\sqrt{2k}}\sum_{\alpha=1,2}\epsilon_{i\vec k \alpha}a_{i \vec k \alpha}A_{i k}(t)e^{i\vec k \cdot \vec r}+h.c.
\end{equation}
where $\epsilon_{i \vec k \alpha}$ are the components of the polarization vectors and $a_{i \vec k \alpha}$ are the usual creator and destruction operators which must satisfy the standard commutation rules:
\begin{align}
&\left[ a_{\perp \vec{k},\alpha},a_{\perp \vec{k'}\beta}^+\right]=\delta_{\vec{k}\vec{k'}}\delta_{\alpha\beta}\\
&\left[ a_{\rVert \vec{k}\alpha},a_{\rVert \vec{k'}\beta}^+\right]=\delta_{\vec{k}\vec{k'}}\delta_{\alpha\beta}\\
&\text{All others}=0
\end{align}
where $\rVert$ means $i=1,2$ and $\perp$ means $i=3$.
It's important to stress the difference by the standard expansion: in this case, it's not possible to define the same modes for all $i$, due to the fact that the equations depend on the particular value of $i$; in fact we obtain:
\begin{align}
\label{eq:perpMode}
&\ddot A_{\perp k}(t)+\left( 2\,\frac{\dot a}{a}-\frac{\dot b}{b}\right)\dot A_{\perp k} (t)+k^2 A_{\perp k}(t)=0\\
\label{eq:parMode}
&\ddot A_{\rVert k}(t)+\frac{\dot b}{b}\,\dot A_{\rVert k} (t)+k^2 A_{\rVert k}(t)=0
\end{align}
that describe the evolution of normal modes coupled to the expanding space-time by $a(t)$ and $b(t)$. In flat FRW case, it is important to note that $a(t)=b(t)$, i.e. Eq.~\eqref{eq:perpMode} and Eq.~\eqref{eq:parMode} became the same equation and so modes evolve at the same manner in all directions. In the statical case $\dot a=\dot b=0$, so Eq.~\eqref{eq:perpMode} and Eq.~\eqref{eq:parMode} admit the well know solution $A_k\sim\exp i\,\omega_k\,t$ with $\omega_k^2=k^2$.

In this Letter we will focus on  a special case. 
Equations \eqref{eq:perpMode} and \eqref{eq:parMode} manifest the presence of several evolving modes: in particular, we can study the high-frequency modes and the low-frequency ones. To this end, let's consider the general equation:
\begin{equation}
\label{eq:gen_mode1}
\ddot A_k(t)+G(t)\dot A_k(t)+k^2A_k(t)=0
\end{equation}
where $G=2\,\frac{\dot a}{a}-\frac{\dot b}{b}$ or $G=\frac{\dot b}{b}$ respectively for transverse and longitudinal modes. At this stage, let's rewrite the Eq.~\eqref{eq:gen_mode1} by introducing $F_k(t)$ as follow:
\begin{equation}
A_k(t)\equiv F_k(t)\exp\left( \pm\,i\,\omega_k\,t\right).
\end{equation}
This is an useful trick that allows us to separate the growth in amplitude from the wave-like evolution of the modes; in fact we obtain the following equation:
\begin{equation}
\label{eq:gen_mode2}
\ddot F_k+\left[ G\pm2\,i\,\omega_k\right] \dot F_k+\left[k^2-\omega_k^2\pm i\,\omega_k\,G\right]F_k=0.
\end{equation}
The behavior of the modes depends on the comparison by $G$ and $\omega_k$: in fact, during $|G(t)|\ll \omega_k$, Eq.~\eqref{eq:gen_mode2} admits as solution:
\begin{equation}
F_k(t)=\left(C_1\,e^{-i\,k\,t}+C_2\,e^{i\,k\,t}\right) \exp{(\mp\,i\,\omega_k\,t)}\Rightarrow A_k(t)=C_1\,e^{-i\,k\,t}+C_2\,e^{i\,k\,t}
\end{equation}
that corresponds to the constant amplitude free-wave modes, with $w_k^2=k^2$. This limit occurs when $\frac{\dot a}{a}\ll \omega_k$ and $\frac{\dot b}{b}\ll \omega_k$, so that we have the very interesting phenomena that the frequency of the oscillation is much higher than the rates of expansion. This means that the modes are able to complete a very large number of oscillations before noticing the change of the space-time that stretches its amplitude.

Otherwise, when $|G(t)|\gg \omega_k$, Eq.~\eqref{eq:gen_mode2} appears as follows:
\begin{equation}
\label{eq:metric_dominated}
\ddot F_k+G(t)\,\dot F_k+[k^2\pm i\,\omega_k\,G(t)]F_k=0.
\end{equation}
This equation is highly non trivial to solve, but we find an heuristic way to understand the behavior of the solutions. Let's suppose that $G(t)\approx q =\text{const}$: in this case Eq.~\eqref{eq:metric_dominated} becomes:
\begin{equation}
\label{eq:euristic_metric_dominated}
\ddot F_k+q\,\dot F_k+[k^2\pm i\,\omega_k\,q]F_k=0 \, ,
\end{equation}
that admits the following solution, assuming $\omega_k\ll q$:
\begin{multline}
F_k\approx \left[ C_1\exp\left[ \left(\frac{k^2}{q}-q\right)t \right] e^{\pm 2\,i\,\omega_k\,t}+C_2\exp\frac{-k^2}{q}t\right] e^{\mp i\,\omega_k\,t}\Rightarrow\\ \Rightarrow A_k\approx C_1\exp\left[ \left(\frac{k^2}{q}-q\right)t \right] e^{\pm 2\,i\,\omega_k\,t}+C_2\exp\frac{-k^2}{q}t.
\end{multline}
Let's stress that the last one is only an heuristic estimation of the right solutions; however  Eq.~\eqref{eq:euristic_metric_dominated} allows some kind of growing amplitude solutions for all modes such that $k>q$. The most general case involves a lot of more complicated situations that allow some mixed cases: for example, it could be happen that $A_\rVert$ is in free wave regime while $A_\perp$ is still in the expanding one (or viceversa).

\section{Revealing interference}
Taking into account all together, it is very interesting to  study the "revealing interference".
Because of the presence of two different kind of evolving modes, there are also two values of energy $E_{\rVert \vec k}$ and $E_{\perp \vec k}$ for a given $\vec k$, corresponding to photon traveling in $x$-$y$ plane and along $z$ axis respectively. In order to consider this feature, we propose the following hamiltonian for system's dynamic, without taking care about polarization's degrees of freedom:
\begin{equation}
\label{eq:anisotropicHamDef}
H\equiv\sum_{\vec k} \left(E_{\rVert \vec k}\,a^+_{\rVert \vec k}\,a_{\rVert \vec k}+E_{\perp \vec k}\,a^+_{\perp \vec k}\,a_{\perp \vec k}\right).
\end{equation}
It is important to underline that Eq.~\eqref{eq:anisotropicHamDef} reduces to the normal ordered hamiltonian of free electromagnetic field in the statical isotropic limit, where $E_{\rVert \vec k}=E_{\perp \vec k}=\hbar\,\omega_k$ and degree of freedom associated to anisotropies vanishes.

At this point, let's define a quantum state for a single photon travelling with momentum $\vec k$ as follow:
\begin{equation}
|\psi>\equiv c_1(\theta)\,a^+_{\perp \vec k}\,|0>+e^{i\alpha}c_2(\theta)\,a^+_{\rVert \vec k}\,|0>
\end{equation}
where the normalization condition must be verified $|c_1|^2+|c_2|^2=1$. Also the following relations must be true: $|c_1(0)|^2=|c_2(\pi/2)|^2=1$ and $|c_1(\pi/2)|^2=|c_2(0)|^2=0$: in this way, a photon traveling along $z$-axis (in $x$-$y$ plane) has energy equal to $E_{\perp \vec k}$ ($E_{\rVert \vec k}$). The simplest choice is $c_1(\theta)=\cos \theta$ and $c_2(\theta)=\sin \theta$. In such a way, 
\begin{equation}
E_\text{one}=\,<\psi|H|\psi>\,=\cos^2 \theta \,E_{\perp \vec k}+\sin^2 \theta \,E_{\rVert \vec k}.
\end{equation}
Now, let's consider a quantum superposition of two photons given by:
\begin{equation}
|\Psi>=\left( \cos \theta\,a^+_{1\perp}+e^{i\alpha}\sin \theta\,a^+_{1\rVert} \right) \left( \cos \phi\,a^+_{2\perp}+e^{i\beta}\sin \phi\,a^+_{2\rVert} \right) |0>
\end{equation}
where $\theta$ and $\phi$ are the angles of the two travel's directions, measured respect to $z$ axis, while $\alpha$ and $\beta$ are the two phases of the photons. In such a way, we are able to calculate the average energy for the given state, i.e.
\begin{figure}[h!]
\centering
\includegraphics[scale=1.2]{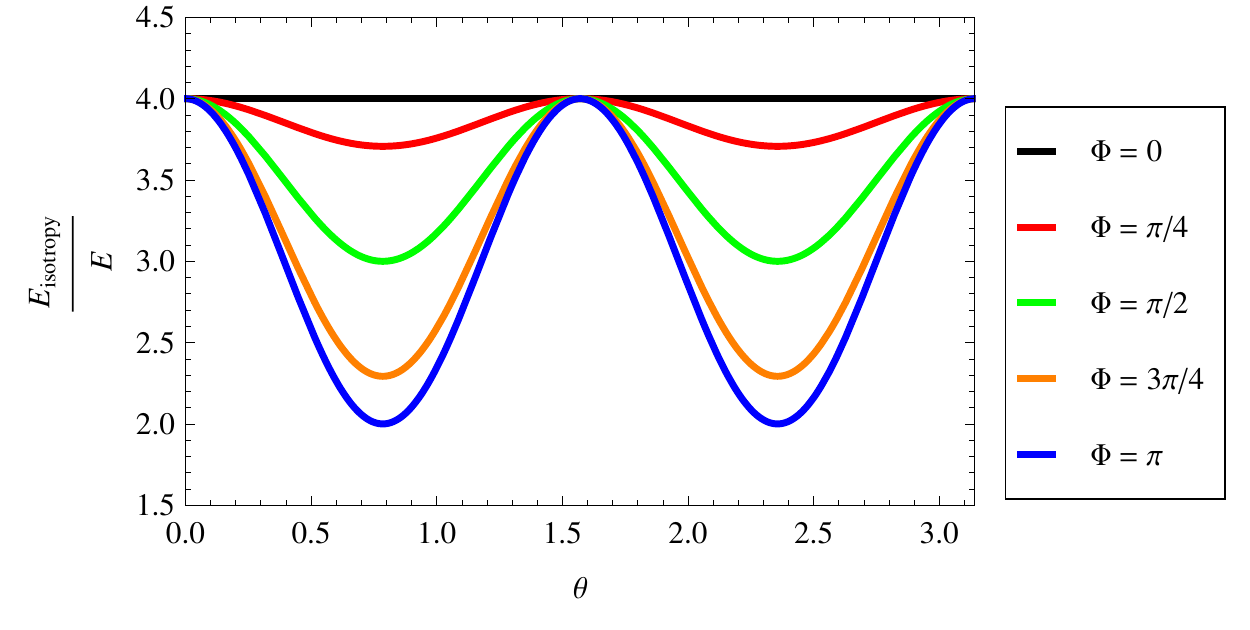}
\caption{Trend of $E_\text{quantum isotropy}$ divided by the energy $E$ of a single photon. Each curve refers to a different relative phase in photons state.}
\label{fig:}
\end{figure}%
\begin{multline}
E_\text{two}=\,<\Psi|H|\Psi>\,=(E_{1\rVert}+E_{2\rVert})\sin^2 \phi \sin^2 \theta+(E_{1\perp}+E_{2\perp})\cos^2 \phi \cos^2 \theta+\\+(E_{1\perp}+E_{2\rVert})\sin^2 \phi \cos^2 \theta+(E_{1\rVert}+E_{2\perp})\sin^2 \theta \cos^2 \phi+2\, \delta_{\vec p_1 \vec p_2}[ E_{1\rVert}\sin^2\phi\sin^2 \theta+E_{1\perp}\cos^2 \theta \cos^2 \phi+\\+(E_{1\rVert}+E_{1\perp}) \sin\phi\sin\theta\cos\phi\cos\theta \cos(\alpha-\beta)].
\end{multline}
In particular, if we consider two photon at the same time traveling along the same direction, i.e. $\theta=\phi$, we obtain:
\begin{multline}
E_\text{same dir}=(E_{1\rVert}+E_{2\rVert})\sin^2\theta+(E_{1\perp}+E_{2\perp})\cos^2\theta+\\+2\,\delta_{\vec p_1\vec p_2}[ E_{1\rVert}\sin^4\theta+E_{1\perp}\cos^4\theta+\sin^2\theta\cos^2\theta (E_{1\rVert}+E_{1\perp})\cos(\alpha-\beta)].
\end{multline}
This is a very interesting expression; in fact, an interference term appears when $\vec p_1=\vec p_2$. In this case, the energies became $E_{1\rVert}=E_{2\rVert} \equiv E_\rVert$ and $E_{1\perp}=E_{2\perp}\equiv E_\perp$ and so we obtain:
\begin{multline}
E_\text{same energy}=2\,E_{\rVert}\sin^2\theta+2\,E_{\perp}\cos^2\theta+\\+2\,[ E_{\rVert}\sin^4\theta+E_{\perp}\cos^4\theta+\sin^2\theta\cos^2\theta (E_{\rVert}+E_{\perp})\cos \Phi]
\end{multline}
where $\Phi\equiv \alpha-\beta$. It's surprising that interference term doesn't disappear when the anisotropy vanishes; in that case, in fact, longitudinal and transverse modes become indistinguishable, i.e. $E_\rVert=E_\perp\equiv E$ so:
\begin{equation}
E_\text{quantum isotropy}=2\,E( 1+\cos^4\theta+\sin^4\theta+2\,\sin^2\theta\cos^2\theta\cos\Phi )
\end{equation}
instead of $E_\text{classical isotropy}= 2\,E$, as expected in the classical case.
This effect appear because of the anisotropyc expansion of the Universe and could constitute early hints for a deviation from the Friedman Lemaitre Robertson Walker metric and the possible existence of a preferred axis in the Universe. 
\\
\section{Conclusion}
This new interesting effect can manifests itself in the electromagnetic emission of far highly directional sources, like quasars. In fact, following \cite{Campanelli:2006vb, Campanelli:2007qn}, we consider a Universe in the matter-dominated era with a plane-symmetric component given by a uniform magnetic field. In the limit of small eccentricity of the Universe we have $e= \sqrt{1 - (b/a)^2}$. We suppose that the magnetic field is frozen it evolves as $B \sim a^{-2}$.
The energy momentum tensor for a uniform magnetic field (in the Universe)  is ${(T_B)}^{\mu}_{\nu} = \rho_B \text{diag} (1,-1,-1,1)$ with $\rho_B = B^2/8 \pi$ the magnetic energy density. In this way
it's possible to obtain the following equation:
\begin{equation}
\frac{d(e\dot e)}{dt}+3 \text{H} (e\dot e)=16\pi G \rho_B
\end{equation}
$\text{H}=\dot a/a$ and $\rho_B$ is the density of a magnetic field that generates the anisotropy. In such a way, we obtain the dependence of eccentricity by the redshift $z$:
\begin{equation}
\label{ez}
e(z)=\left(\frac{z}{z_\text{dec}}\right)^{3/4} e_\text{dec}.
\end{equation}
Remebering that $e_\text{dec}\sim 10^{-2}$ and $z_\text{dec}=1088$ are valued at the decoupling era, we obtain that the eccentricity is $\sim 10^{-4}$ for the farest observed quasars ($z=7$). In other words the analysis of the spectra of the far quasars may be a good check to study the anisotropy of the Universe. On the other hand it is possible that we have a local effect of anisotropic expansion in a cosmological isotropic background, in this way the local anisotropic expansion may be localized by quasars that may show the eccentricity of Eq.(\ref{ez}). This requires a detailed study that we postpone to future analysis.  
\\
\\
\section*{Aknowledgements}

This work is supported by the research grant Theoretical Astroparticle Physics No. 2012CPPYP7 under the program PRIN 2012 funded by the Ministero dell'Istruzione, Universit\'a e della Ricerca (MIUR). This work is also supported by the italian istituto nazionale di fisica nucleare (infn) through the theoretical astroparticle physics project.
%******************************************************

 %
 \end{document}